\documentclass[preprint,showpacs,preprintnumbers,amsmath,amssymb,nofootinbib,showkeys,aps,usenatbib]{revtex4}
\pdfoutput=1
\usepackage{epsfig}
\usepackage{mathtools}
\usepackage{multirow}
\usepackage{graphicx}% Include figure files
\usepackage{dcolumn}% Align table columns on decimal point
\usepackage{bm}% bold math
\usepackage{amsmath}
\usepackage{amssymb}
\usepackage{latexsym}
\usepackage{color}
\usepackage{hyperref}
\usepackage{mathrsfs}
\usepackage[amsmath]{empheq}
\usepackage{float} 

\newcommand{\be}{\begin{equation}}
\newcommand{\ee}{\end{equation}}
\newcommand{\bea}{\begin{eqnarray}}
\newcommand{\eea}{\end{eqnarray}}
\newcommand{\texpdf}{\texorpdfstring}

\begin{document}

%\preprint{APS/123-QED}

\title{Gaussian Processes Reconstruction of the Dark Energy Potential}

\author{J. F. Jesus$^{1,2}$}\email{jf.jesus@unesp.br}
\author{R. Valentim$^{3}$} \email{valentim.rodolfo@unifesp.br}
\author{A. A. Escobal$^{2}$} \email{anderson.aescobal@gmail.com}
\author{S. H. Pereira$^{2}$} \email{s.pereira@unesp.br}
\author{D. Benndorf$^{2}$} \email{douglas.benndorf@unesp.br}

\affiliation{$^1$Instituto de Ciências e Engenharia, Universidade Estadual Paulista (UNESP) - R. Geraldo Alckmin, 519, 18409-010, Itapeva, SP, Brazil,\\
\\$^2$Departamento de F\'isica, Faculdade de Engenharia de Guaratinguet\'a, Universidade Estadual Paulista (UNESP) - Av. Dr. Ariberto Pereira da Cunha 333, 12516-410, Guaratinguet\'a, SP, Brazil,\\
\\$^3$Departamento de F\'{\i}sica, Instituto de Ci\^encias Ambientais, Qu\'{\i}micas e Farmac\^euticas (ICAQF), Universidade Federal de S\~ao Paulo (UNIFESP) - Rua S\~ao Nicolau 210, 09913-030, Diadema, SP, Brazil.}

%  but any date may be explicitly specifie

%%%%%%%%%%%%%%%%%%%%%%%%%%%% Include extra macros I used (MSSG)
\def\zt{\mbox{$z_t$}}

%\vspace{0.5cm}
\begin{abstract}
%\begin{center}{\bf ABSTRACT}\end{center}

Scalar Fields (SF) have emerged as natural candidates for dark energy as quintessential or phantom fields, as they are the main ingredient of inflation theories. Instead of assuming some form for the scalar field potential, however, this work reconstructs the SF potential directly from observational data, namely, {Hubble and SNe Ia data}. We show that two popular forms for the SF potentials, namely, the power-law and the quadratic free-field, are compatible with the reconstructions thus obtained, at least for some choices of the priors of the matter density and curvature parameters and for some redshift intervals.
%\vspace{0.1cm}
\end{abstract}

%\pacs{98.80-k; 98.80.Es; 98.80.Cq}
\maketitle

%\newpage

%\vspace{0.1cm}

\section{Introduction} 

The standard model of cosmology, namely flat $\Lambda$CDM model, predicts the current acceleration for the universe and it has been consistent with many observations \cite{Farooq:2016zwm,Scolnic:2017caz,Aghanim:2018eyx}, although there exist some theoretical and observational discrepancies \citep{Riess:2020sih,Martinelli:2019krf,Bull:2015stt} that it opens the possibility for other cosmological models, as scalar field quintessence or dark energy models \cite{Sahni:1999qe,Rubano:2001xi,Paliathanasis:2014zxa,Dimakis:2016mip,Paliathanasis:2015cza,Paliathanasis:2014yfa,Barrow:1993ah,Lee:2004vm}. Unified cosmological models for inflation, dark matter and dark energy have been constructed with scalar fields have also proposed recently \cite{Sa:2020fvn,Lin:2009ta,Liddle:2008bm} and it constraints on quintessence scalar field models using cosmological observations have also been  carried out \cite{Yang:2018xah,Cao:2020jgu,Urena-Lopez:2020npg,Khadka:2019njj,Singh:2018izf,Sola:2017jbl}. The selection of a quintessence scalar field model can be done by choosing an appropriate potential $V(\phi)$ to drive different dynamical phases of the Universe, and its free parameters must be constrained in order to be in agreement with observational data. Since the study of a simple quadratic free-field potential \cite{Ratra:1990me,Urena-Lopez:2007zal} to power-law
type potential was proposed by Peebles and Ratra \cite{Peebles:1987ek,Ratra:1987rm}, several potentials have been studied on the context of dark energy models.  Recently, one work has proposed by Yang et al. \cite{Yang:2018xah} and the authors consider some very general potentials, including exponential, hyperbolic and general power law type, where the potentials were studied in order to constraint the model with baryon
acoustic oscillations (BAO), the cosmic microwave background observations (CMB),  joint light curve analysis (JLA) from supernovae type Ia (SNe Ia), redshift space
distortions (RSD) and the cosmic chronometers (CC). Recently, the power-law potential was used to constraint cosmological parameters from HII starburst galaxy apparent magnitude plus other cosmological measurements \cite{Cao:2020jgu}.

The above mentioned potentials for quintessence or dark energy models are just few examples of several potentials already studied in the literature. However, as it was recently showed, the potentials should not be much arbitrary, then it must have some constraints according to swampland criteria \citep{Obied:2018sgi,Agrawal:2018own,Heisenberg:2018yae,Heisenberg:2018rdu}. Such criteria imposes restrictive features on quintessence scalar field models, such as that the
quintessence potential should not be steeper than order unity in Planck units. Another way to find a suitable potential is by reconstructing it directly from observational data. 

An interesting method to reconstruct cosmological parameters of a model has been proposed recently by Seikel, Clarkson and Smith \cite{Seikel:2012uu} by using Gaussian Process (GP). Statistical tool approach is based on a non-parametric method to reconstruct the dependence with redshift of a cosmological observable, as luminosity distance \cite{Seikel:2012uu}, equation of state parameter of dark energy \citep{Holsclaw:2010nb}, the Hubble parameter \cite{Shafieloo:2012ht} or transition redshift for instance \cite{Jesus:2019nnk}. With the GP method, one can reconstruct a general function of the model directly from data,
without the need of a particular parameterization for it. The reconstruction of a DE scalar field potential was done in \cite{Li:2006ea,Sahlen:2005zw} and limits from SNe Ia data was obtained recently by \cite{Piloyan:2018xwa}. There are several other recent analysis of cosmological parameters  reconstructed through GP method. In \cite{Bernardo:2021mfs} a model-independent reconstruction approach for late-time Hubble data was done. In \cite{Krishak:2021fxp} the reconstruction of the reionization history was analysed. In \cite{Escamilla-Rivera:2021rbe} the effectiveness of non-parametric reconstruction techniques was used in the context of the Hubble tension problem. In \cite{Sun:2021pbu} the influence of the bounds of the hyperparameters on the reconstruction of Hubble constant with GP was studied. The nonparametric reconstruction of interaction in the cosmic dark sector was done in \cite{Mukherjee:2021ggf,vonMarttens:2020apn}. The cosmic distance duality relation using GP was analysed in \cite{Mukherjee:2021kcu}.

In the present paper the main aim to reconstruct the dark energy scalar field potential directly from $H(z)$ and SNe Ia data, taking into account the spatial curvature.

The paper is organized as follows. In section \ref{dyn} the main equations are presented. In section \ref{datamet} the dataset and Gaussian Process methodology are described by us. In section \ref{results} the results are showed. Conclusions are left to section \ref{conc} and in the Appendix \ref{app} we show the equations for the particular case of spatial flatness.

\section{\label{dyn}Cosmological Dynamics}

For a quintessence  cosmological model with  curvature, the Friedmann equations can be written as
\begin{align}\label{EF}
H^2 &= \frac{8\pi G}{3}(\rho_m + \rho_\phi)-\frac{k}{a^2}\,,\\\label{EA}
\frac{\ddot a}{a} &= -\frac{4\pi G}{3}\left(\rho_m +\rho_\phi + 3p_\phi\right)\,,
\end{align}
where $\rho_{m}$ and $p_m$ are the energy density and pressure of matter content and $\rho_{\phi}$ and $p_{\phi}$ are  the energy density and pressure of the scalar field $\phi$, given by
\begin{align}\label{DEP}
    \rho_{\phi} = \dfrac{1}{2}\dot{\phi}^2+V(\phi),\,\,\,p_{\phi} = \dfrac{1}{2}\dot{\phi}^2-V(\phi)\,.
\end{align}
The potential $V(\phi)$ carries the information about the time evolution of the homogeneous scalar field. We can rewrite the system \eqref{EF}-\eqref{EA} as
\begin{align}\label{EF2}
    H^2 &= \frac{8\pi G}{3}\left[\rho_m +\frac{\dot\phi^2}{2}+V(\phi)\right]-\frac{k}{a^2}\,,\\\label{EA2}
\frac{\ddot a}{a} &= -\frac{4\pi G}{3}\left[\rho_m +2
\dot\phi^2-2V(\phi)\right]\,.
\end{align}
By eliminating $\dot{\phi}^2$ in \eqref{EF2}-\eqref{EA2} and using the relation $\frac{\ddot{a}}{a}=\dot{H}+H^2$, we may express the potential $V(\phi)$ as
\be\label{EP}
V(\phi) = \frac{3H^2+\dot H}{8\pi G}-\frac{\rho_m}{2}+\frac{k}{4\pi Ga^2}\,,
\ee 
and for a numerical analysis, it is useful to use the dimensionless potential $U(\phi)$ defined as
\begin{align}\label{PA}
    U(\phi)\equiv\frac{8\pi G}{3H_0^2}V(\phi).
\end{align}

The observational data used in potential reconstruction are given in terms of {\it redshift} $z$. Thus, it is necessary to change the temporal dependence of the equation \eqref{EP} to a dependence on $z$ through the relation
%Now we make a change of variables:
\be\label{TZ}
\frac{d}{dt}=-H(1+z)\frac{d}{dz}\,.
\ee
Using the definition \eqref{PA} and the relation \eqref{TZ}, the equation for the potential $V(\phi)$ given by \eqref{EP} is now written as
%So:
\be
U(z)=E^2-\frac{E(1+z)}{3}\frac{dE}{dz}-\frac{\Omega_{m}(1+z)^3}{2}-\frac{2\Omega_{k}}{3}(1+z)^2\,.
\label{UphiEz}
\ee
where $E(z)\equiv\frac{H(z)}{H_0}$, $\Omega_{m}\equiv\frac{8\pi G\rho_{m0}}{3H_0^2}$ and $\Omega_{k}\equiv-\frac{k}{a_0^2H_0^2}$ are dimensionless quantities related to Hubble parameter, matter density and curvature parameter, respectively. We have also used the fact that the pressureless matter is separately conserved, so that $\frac{8\pi G\rho_m}{3H_0^2}=\Omega_{m}(1+z)^3$.

{On the other hand, by eliminating $V(\phi)$ in \eqref{EF2}-\eqref{EA2}, we may write for the kinetic energy of the scalar field $T\equiv\frac{\dot{\phi}^2}{2}$:}
\be
T=-\frac{\dot{H}}{8\pi G}-\frac{\rho_m}{2}+\frac{k}{8\pi Ga^2}
\ee
{from which we can define the dimensionless kinetic energy $\tau\equiv\frac{8\pi G}{3H_0^2}T$ and write it in terms of the redshift as:}
\be
\tau(z)=\frac{E(1+z)}{3}\frac{dE}{dz}-\frac{\Omega_{m}(1+z)^3}{2}-\frac{\Omega_{k}}{3}(1+z)^2\,.
\label{tauz}
\ee

{The expressions \eqref{UphiEz} and \eqref{tauz} for the dimensionless potential $U(z)$ and kinetic energy $\tau(z)$ are written as functions of the {\it redshift} and other dimensionless quantities, in particular, $E(z)$. This allows us to reconstruct $U(z)$ and $\tau(z)$ by reconstructing $E(z)$ and its derivatives via GP using the data from $H(z)$.
Similarly, to use the reconstruction made by the GPs for the SNe Ia data, it is advantageous to express the scalar field potential $U(z)$ and kinetic energy $\tau(z)$ in terms of the dimensionless transverse comoving distance $D_M(z)$, which is given by}
\be
D_M(z)=\frac{1}{\sqrt{-\Omega_k}}\sin\left(\sqrt{-\Omega_k}D_C(z)\right)
\label{DMDC}
\ee
where the line-of-sight comoving distance $D_C(z)$ relates to $E(z)$ as:
\be
D_C'(z)= \frac{1}{E(z)}\,,
\label{DCEz}
\ee
where a prime denote a derivative with respect to redshift $z$. Dimensionless distances $D_i$ relate to dimensionful distances $d_i$ as:
\be
D_i\equiv\frac{d_i}{d_H},
\ee
where $d_H\equiv\frac{c}{H_0}$ is Hubble distance.

The derivative with respect to $z$ of the transverse comoving distance $D_M(z)$ is
%From \eqref{DMDC}, we find:
\be
\dfrac{d D_M(z)}{dz}=D_M'(z)=\cos\left(\sqrt{-\Omega_k}D_C(z)\right)D_C'(z)\,,
\label{dDMDC}
\ee
so, by combining \eqref{DMDC} and \eqref{dDMDC}, we find:
\be\label{EDM}
\left(\frac{D_M'}{D_C'}\right)^2-\Omega_kD_M^2=1\,.
\ee
Now, by using \eqref{DCEz}, we can express the relation \eqref{EDM} in terms of the dimensionless quantity $E(z)$:
%Using \eqref{DCEz}:
\be\label{EFDM}
E^2=\frac{1+\Omega_kD_M^2}{D_M'^2}\,.
\ee
Taking the derivative with respect to $z$ of the relation \eqref{EFDM}, we have
\be
E\frac{dE}{dz}=\frac{\Omega_kD_M\left(D_M'^2-D_MD_M''\right)-D_M''}{D_M'^3}\,,
\ee
which allows us to write the dimensionless potential $U(\phi)$ given by \eqref{UphiEz} as a function of the transverse comoving distance $D_M(z)$ and its derivatives, $D'_M(z)$ and $ D''_M(z)$:
\be
U(z)=\frac{1+\Omega_kD_M^2}{D_M'^2}+\left(\frac{1+z}{3}\right)\frac{D_M''+\Omega_kD_M\left(D_MD_M''-D_M'^2\right)}{D_M'^3}-\frac{\Omega_{m}(1+z)^3}{2}-\frac{2\Omega_{k}}{3}(1+z)^2\,. \label{UzDM}
\ee

{Similarly, the dimensionless kinetic energy \eqref{tauz} can be obtained from the transverse comoving distance as:}
\be
\tau(z)=\left(\frac{1+z}{3}\right)\frac{\Omega_kD_M\left(D_M'^2-D_MD_M''\right)-D_M''}{D_M'^3}-\frac{\Omega_{m}(1+z)^3}{2}-\frac{\Omega_{k}}{3}(1+z)^2\,. \label{tauzDM}
\ee

{In the present analyses, we have assumed that $\tau>0$ from both reconstructions, from Eqs. \eqref{tauz} and \eqref{tauzDM}}.

\section{\label{datamet}Dataset and methodology}
The observational data sample used in this work includes a compilation of measurements of the Hubble parameter, $H(z)$ \cite{Magana:2017nfs}, and a large dataset of SNe Ia  from the Pantheon compilation \cite{Scolnic:2017caz}.

In order to maintain the analysis cosmological model independent, we shall focus on the $31$ astrophysical measurements of $H(z)$, distributed over the range $0.07 <z< 1.965$, obtained through the estimate of differential ages of galaxies \cite{Simon2004, Stern2009, Moresco2012, Zhang2012, Moresco2015, Moresco2016}, called cosmic chronometers.

%In order to maintain the cosmological model independence of the analysis, we shall focus on the 31 cosmic chronometers $H(z)$ data.

The SNe Ia data, on the other hand, consist of the Pantheon sample, which has $1048$ measurements of SNe Ia, in the redshift range $0.01\,<\, z\,<\,2.3$, containing measurements from Pan-STARRS1 (PS1), SDSS, SNLS, and various low-$z$ and HST datasets .

%%% GP

From the data set described above, we will use a nonparametric method called Gaussian Processes (GP) \cite{Seikel:2012uu}, which allows the reconstruction of a continuous function, $f(x)$, and its derivatives through the discrete set of values of this function, in which each of the values is assumed to represent a random variable that follows a Gaussian distribution. As this set of points correspond to a same underlying function to be reconstructed,
%When trying do make this reconstruction, we must take into account the correlation between points of the function to be reconstructed.
%the error due to the dependence of the distance of the function values between each point it is being evaluated. Then, 
it is necessary to choose a correlation function $k(x_i,x_j)$ between the points $x_i$ and $x_j$. There are several correlation functions available in the literature \cite{williams2006gaussian,Seikel:2012uu,Jesus:2019nnk}. {As shown in many references, for instance, \cite{Jesus:2019nnk,YuEtAl17}, there is no relevant difference in the reconstructions while choosing among the most commmon correlation functions (kernels). So, that is why we choose to work here with the most common kernel, that is called Squared Exponential}, given by
\begin{align}
k(x_i,x_j)=\sigma_f^2\exp\left[-\frac{(x_i-x_j)^2}{2l^2}\right]\,,
\end{align}
where $l$ and $\sigma_f$ are the so-called GP hyperparameters that must be determined from the data.

With the 31 cosmic chronometers data, we can directly reconstruct the function $H(z)$ and its derivatives. By using the relationship \eqref{UphiEz} we can reconstruct the dimensionless potential $U(z)$ associated with the scalar field $\phi$. The transverse comoving distance $D_M$ and its derivatives can be reconstructed via GP using the Pantheon SNe Ia apparent magnitudes $m_B$ and, together with the expression \eqref{UzDM}, we can perform the reconstruction of the dimensionless potential $U(z)$ from Pantheon data.

We used the GaPP software \cite{Seikel:2012uu} in order to implement the Gaussian Processes to reconstruct $H(z)$, $D_M(z)$ and their derivatives. Then, we developed a parallelized code to find $U(z)$ and $D_M(z)$ by sampling the multivariate Gaussian distributions involving $H(z)$ and its derivatives and $D_M(z)$ and its derivatives.

%%%%%%%%%%%%%%%%%%% Potenciais de teste
The specific form of the potential $V(\phi)$, in addition to determining the temporal evolution of the field $\phi$, can provide important information about the quintessence model, such as the mass associated with the DE particle and the self-interaction terms. We will make a comparison of the potential $U(z)$ reconstructed in our method with some already standard models of scalar field dark energy present in the literature. We will focus on the old quadratic free-field potential, $U(\Phi)_{FF}$ \cite{Ratra:1990me,Urena-Lopez:2007zal} and the Peebles-Ratra power law potential \cite{Peebles:1987ek,Ratra:1987rm}, $U(\Phi)_{PL}$. The quadratic free-field potential is written as
\begin{align}
     U(\Phi)_{FF}    &= \dfrac{\mu^2 }{2}\Phi^2\,,
\end{align}
where the parameter $\mu\equiv m/H_0$ is related to the mass $m$ of the scalar field and $\Phi\equiv\sqrt{\frac{8\pi G}{3}}\phi$ is the dimensionless scalar field, while the Power Law potential is given by
\begin{align} 
    U(\Phi)_{PL} &= \dfrac{\kappa  \Phi^{-\alpha}}{2}\,
\end{align}
where $\kappa$ is a constant that is given in terms of the coefficient $\alpha$ as
\begin{align}
    \kappa = \dfrac{8}{3}\left( \dfrac{\alpha+4}{\alpha+2}\right)\left[ \dfrac{2}{3}\alpha\left(\alpha+2\right)\right]^{\alpha/2}\,.
\end{align}
When $\alpha = 0$ this potential reduces to the standard $\Lambda$CDM model.

\section{\label{results}Results}
Using the GP method, as described in the previous section, we were able to reconstruct the functions $H(z)$ and $D_M(z)$ using data from the Hubble parameter and the Pantheon sample, respectively. However, in order to obtain the dimensionless potential $U(z)$ from these reconstructions, as one can see from Eqs. \eqref{UphiEz} and \eqref{UzDM}, there are free parameters that should be obtained, namely, $\Omega_m$ and $\Omega_k$. These parameters can not be obtained from the reconstruction and should be constrained from other observations instead, which can furnish priors over these parameters. Once we have priors over $(\Omega_m,\Omega_k)$, we can obtain $U(z)$ by sampling these priors and the multivariate Gaussian corresponding to $(H(z),D_M(z))$ and their derivatives.

{Besides the reconstructions from $H(z)$ data and Pantheon sample, we have also made a joint analysis in $U(z)$ reconstruction, where there was 1$\sigma$ compatibility between both reconstructions. In order to do that, we have weighted the SNe Ia reconstruction with the $H(z)$ reconstruction, which is a method of combining MCMC chains, known as importance sampling.}

We chose to work with 2 different priors in order to obtain robust results. The first prior analyzed was that of Planck 18 \cite{Aghanim:2018eyx} $(\Omega_k = -0.044\pm0.050, \Omega_m=0.315\pm0.022)$ at 3$\sigma$ c.l. We chose to work with 3$\sigma$ c.l. priors, as these Planck 18 results do not correspond to scalar field dark energy cosmologies, but to the $\Lambda$CDM model.

The second prior we have used was a ``large'' prior on the parameters, which encompass most of the current observations, namely, $(\Omega_k = 0.0\pm0.1, \Omega_m=0.30\pm0.05)$.

{In our analyses, we have also assumed $\tau>0$, which corresponds to $\dot{\phi}^2>0$, which is a physical condition necessary for the validity of the class of scalar field models. In order to do that, when obtaining the $U(z)$ reconstructions, we have neglected the samples with $\tau<0$.}

\begin{figure}
    \centering
    \includegraphics[width=.49\textwidth]{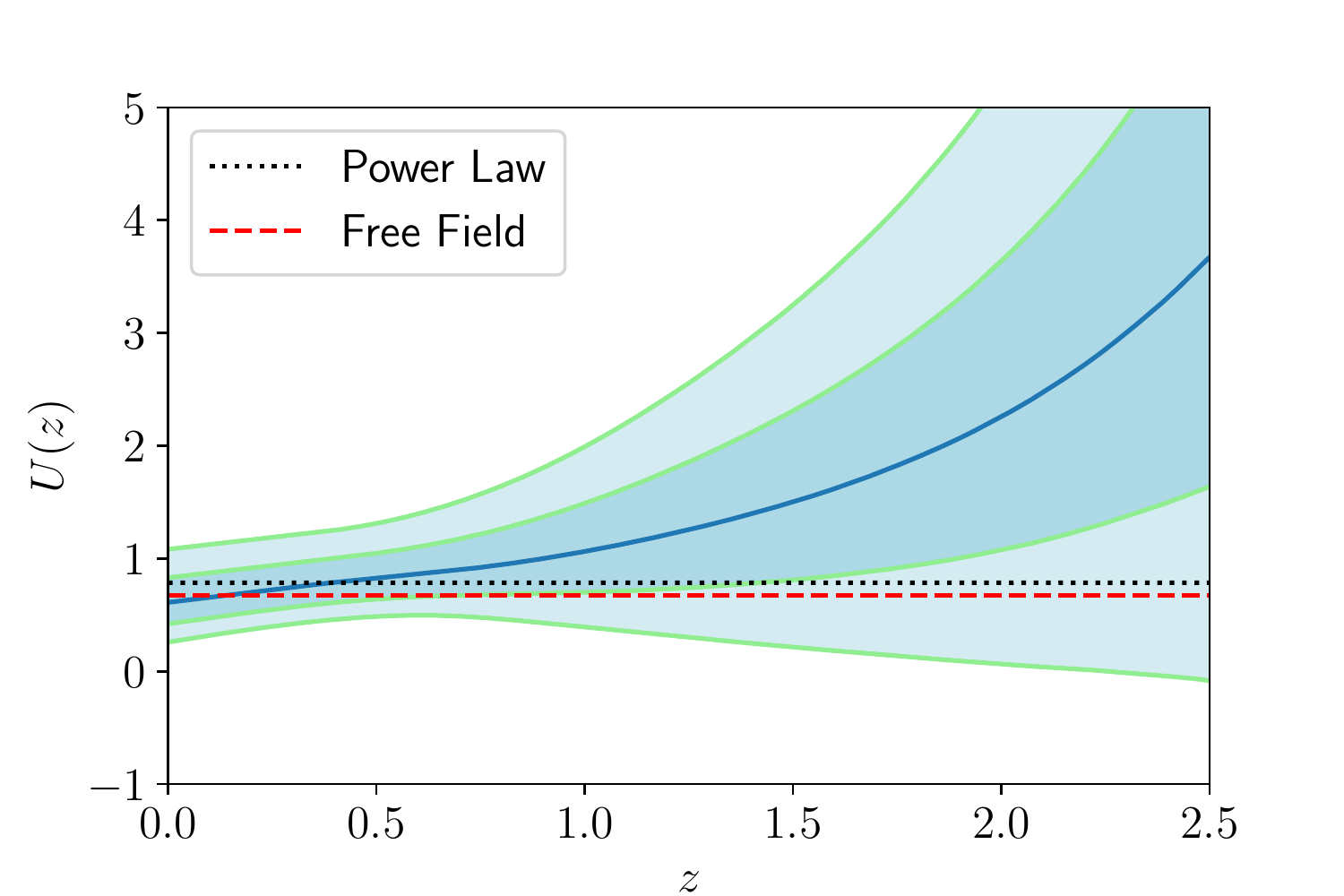}
    \includegraphics[width=.49\textwidth]{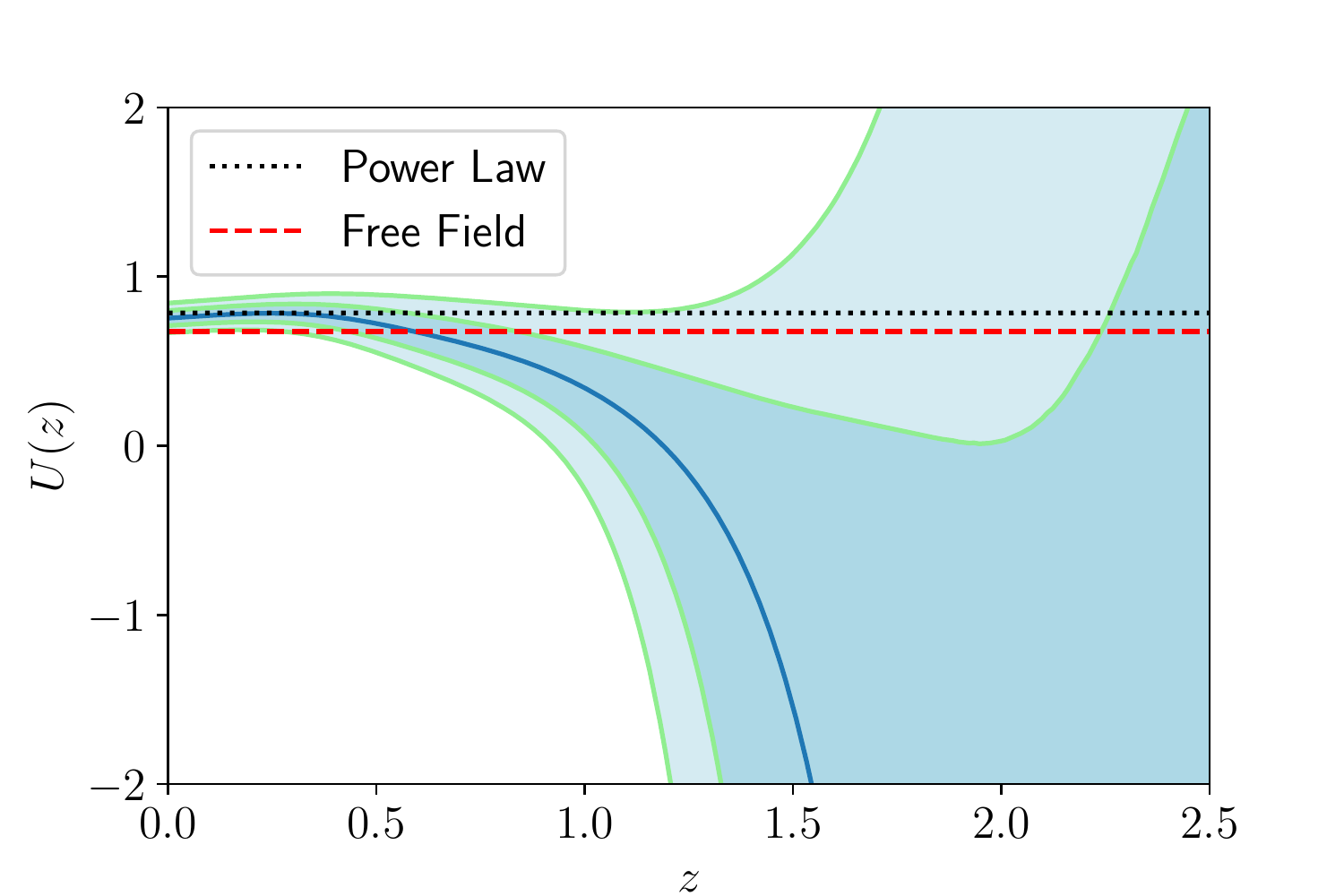}
    \caption{Results from the reconstructions with the Planck 3$\sigma$ prior. {Left:} $U(z)$ reconstruction from 31 $H(z)$ data.  {Right:} $U(z)$ reconstruction from Pantheon.}
    \label{PRPP}
\end{figure}

\begin{figure}
    \centering
    \includegraphics[width=.49\textwidth]{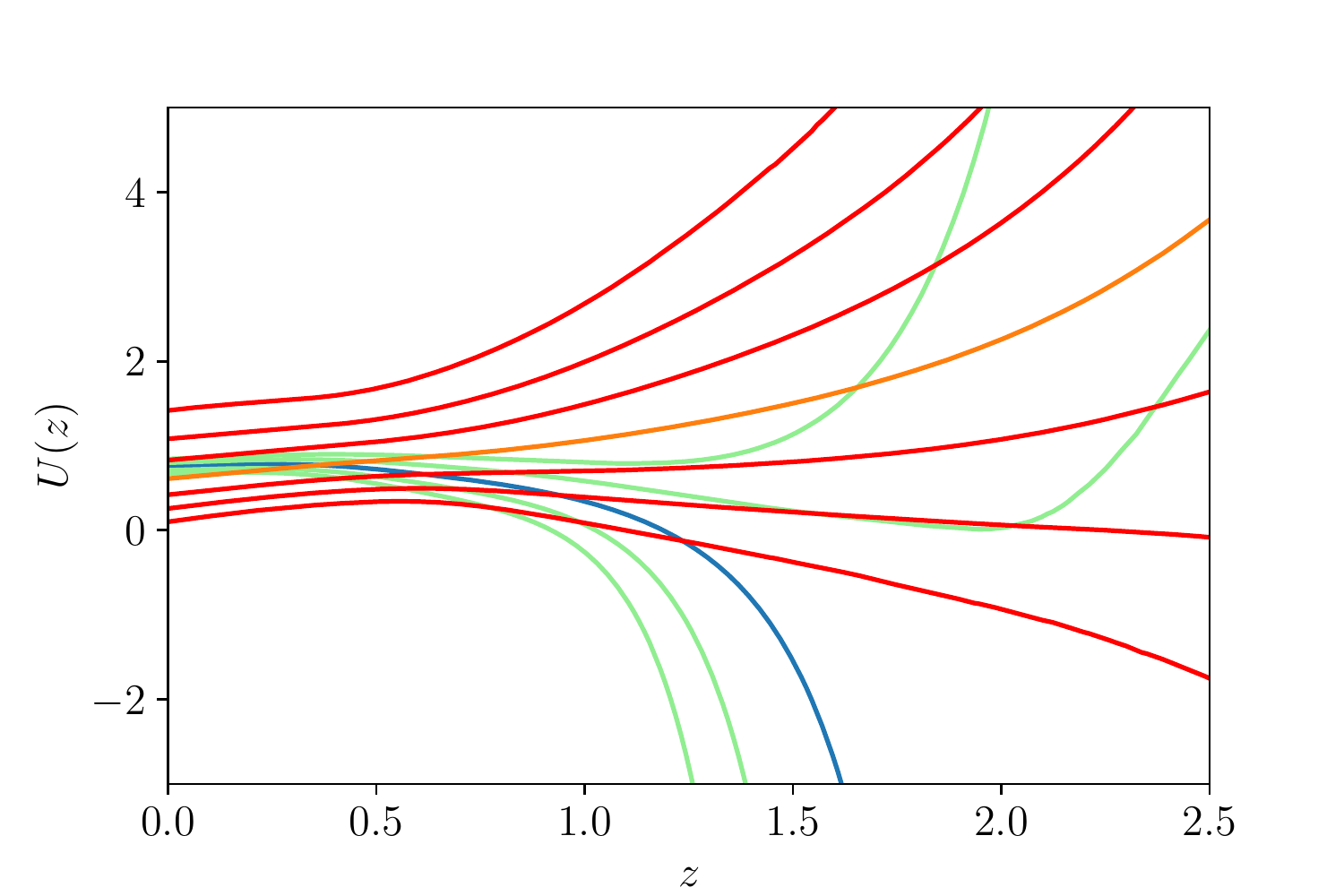}
    \includegraphics[width=.49\textwidth]{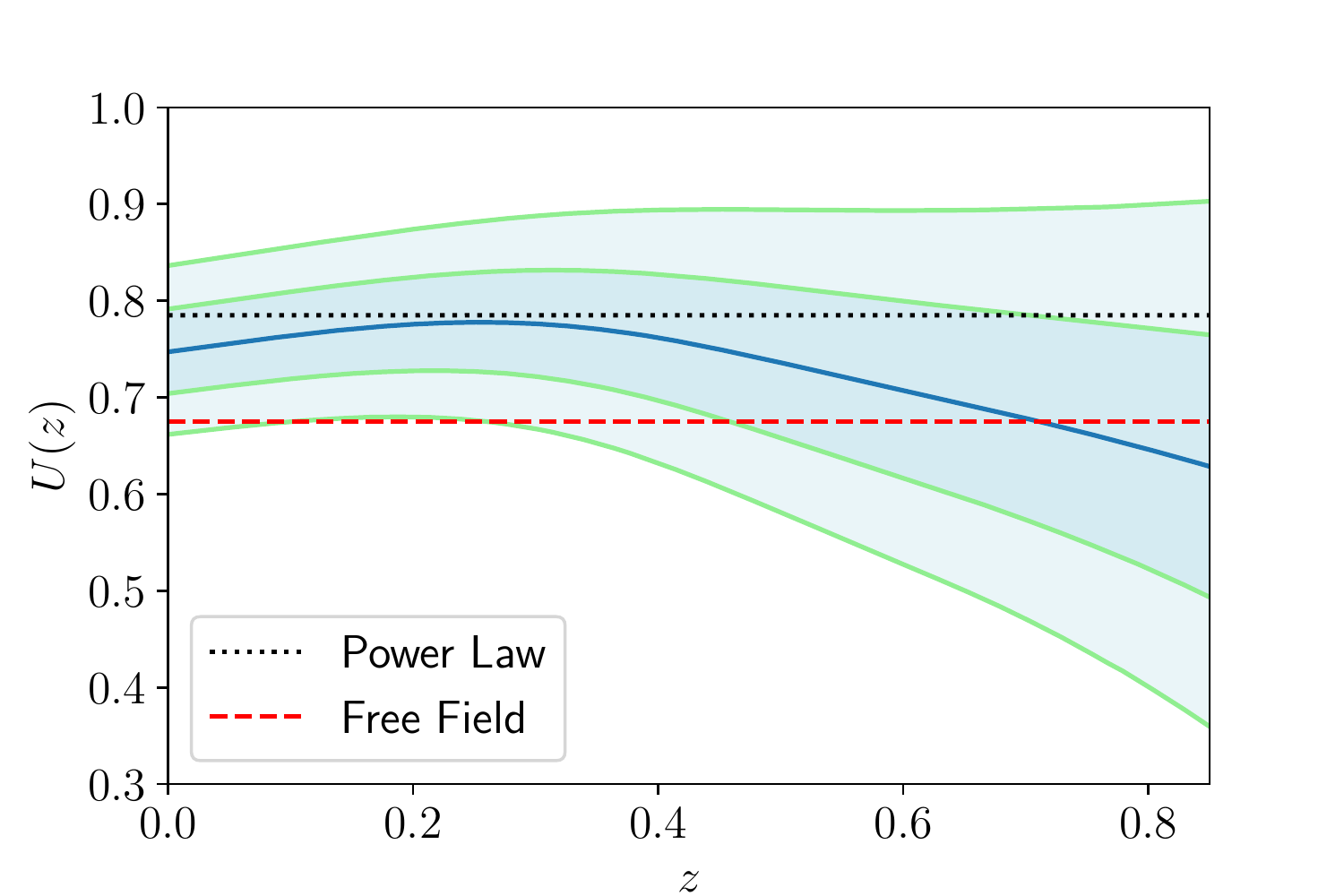}
    \caption{$U(z)$ reconstruction from $H(z)$ and Pantheon data with a Planck 3$\sigma$ prior over $\Omega_m$ and $\Omega_k$. {Left:} Superposition of confidence intervals. SNe Ia reconstruction corresponds to red and $H(z)$ reconstruction corresponds to green continuous lines. {Right:} Joint analysis.}
    \label{PCPP}
\end{figure}

{The results obtained using PP, are} presented in Figs. \ref{PRPP} and \ref{PCPP}. In Fig. \ref{PRPP}, we show the reconstruction of $U(z)$ within a $2\sigma$ confidence interval, from $H(z)$ data (left), and from Pantheon SNe Ia data (right). {In both cases, we also show the evolution of the dimensionless potential for two test models: the Power Law (PL) model, $U_{PL}(z)$, and the quadratic free-field (FF) model, $U_ {FF}(z)$. In the case of PL, we have assumed $\alpha = 0.150$ from the analysis performed by \cite{Cao:2020jgu}. For FF, we assumed the mass $m = 0.60 H_0$ eV, which we obtained through statistical constraints from the $H(z)$ data and Pantheon sample.} We can observe that the reconstruction from $H(z)$ data has a better compatibility with the curves of $U_{PL}$ and $U_{FF}$, in comparison with the SNe Ia reconstruction. {The potential $U_{PL}(z)$, in particular, is compatible with rebuilding $U(z)$ with the data of $H(z)$ at less than $1\sigma$ at nearly $60\%$ of the entire analyzed redshift range, while the potential $U(z)_{ FF}$ is between the lines $1\sigma$ and 2$\sigma$ from $z\gtrsim 0.72 $.}

On the other hand, the reconstruction of $U(z)$ obtained from Pantheon data, has a small uncertainty in the $z<1$ range, causing the contour regions up to $2\sigma$ to be well constrained. In this case, in the remainder of the redshift interval, the width of the $1\sigma$ confidence region increases according to the form that the $U(z)$ reconstruction assumes, and this leads to less compatibility among the reconstruction and both potentials $U_{FF}(z)$ and $U_{PL}(z)$.{We see that the $U_{PL}(z)$ curve is in the region of $1\sigma$ of the reconstruction up to $z\sim 0.58$ and, at $z\gtrsim 2.2$, the curve becomes compatible again in less than 1$\sigma$ of the reconstruction of $U(z)$. The $U_{FF}(z)$ curve is compatible at less than 1$\sigma$ with reconstruction in approximately $25\%$ of the redshift range, however it is slightly lower than the $2\sigma$ reconstruction line in the range from $0.05\lesssim z\lesssim 0.28$.}

In Fig. \ref{PCPP} (Left), we have the superposed plot of the $U(z)$ reconstructions for both observational data samples used, which were plotted with a confidence interval of 1 and 2$\sigma$. We can see that for $z\lesssim0.5$, the $2\sigma$ reconstruction of $U(z)$ for the SNe Ia data is within the range of $1\sigma$ of the reconstruction of $U(z)$ with the data of $H(z)$.  {The two reconstructions show only a small divergence for $z\gtrsim 0.85$, as there is no overlap of their 1$\sigma$ confidence regions. In this interval, $0 \le z <0.85$, with compatibility of $1\sigma$ between the reconstructions, we have made a joint analysis between $H(z)$ data and Pantheon sample, in order to obtain a joint reconstruction for $U(z)$, as shown in the Figure \eqref{PCPP} (Right). The $U(z)_{PL}$ curve remains in the region of $1\sigma$ up to $z \sim 0.7$, while the $U(z)_{FF}$ curve
was below the region of $2\sigma$ in the range $0.10 < z < 0.26$ and only for $z\gtrsim 0.46$ compatibility region reaches $1\sigma$ with the reconstruction of $U(z)$ from the data sample combination.}
%Furthermore, the biggest divergence between the reconstructions occurs due to the decrease in the potential obtained from the SNe Ia data, since the reconstruction from the $H(z)$ data is more ``well-behaved''.

\begin{figure}
    \centering
    \includegraphics[width=.49\textwidth]{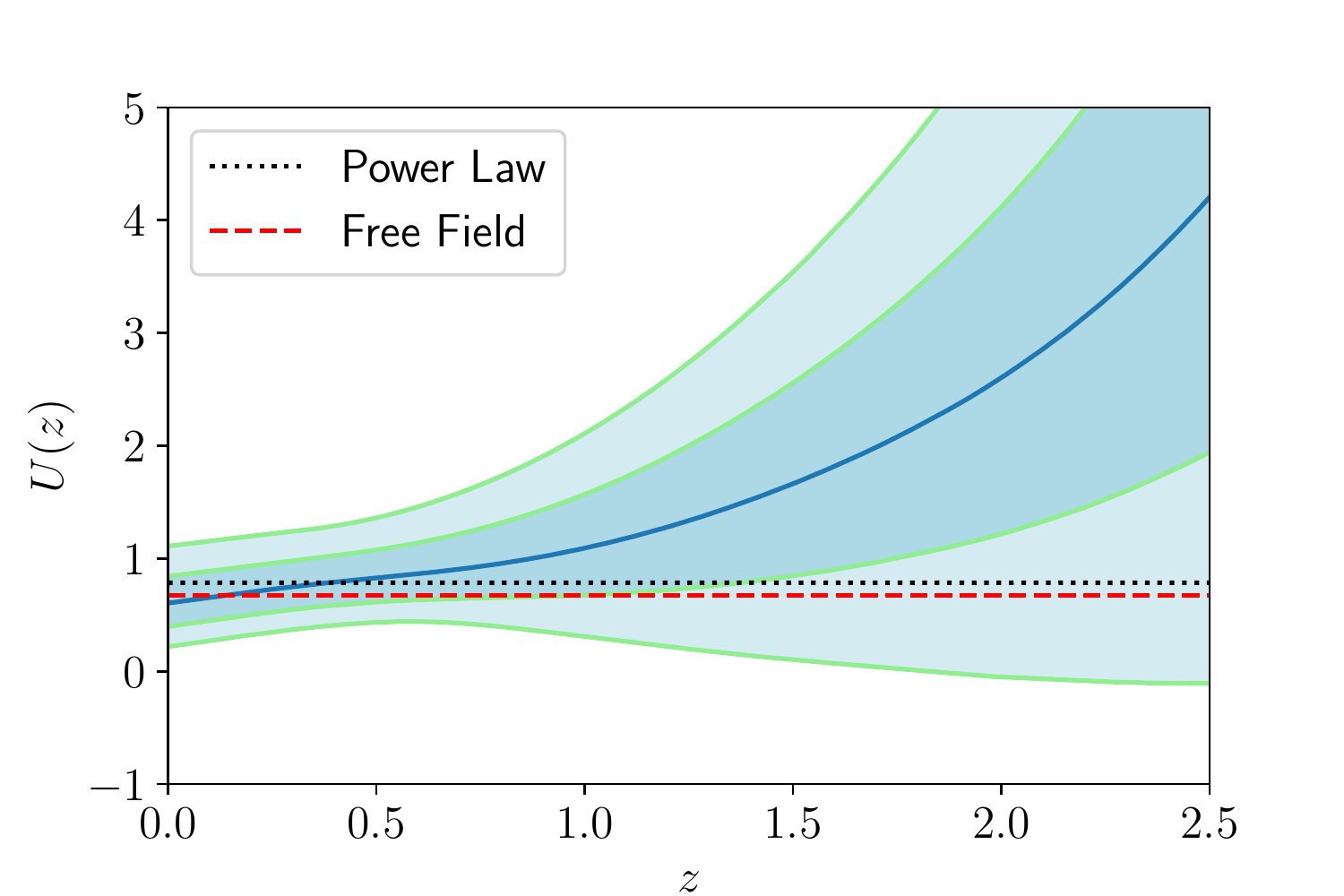}
    \includegraphics[width=.49\textwidth]{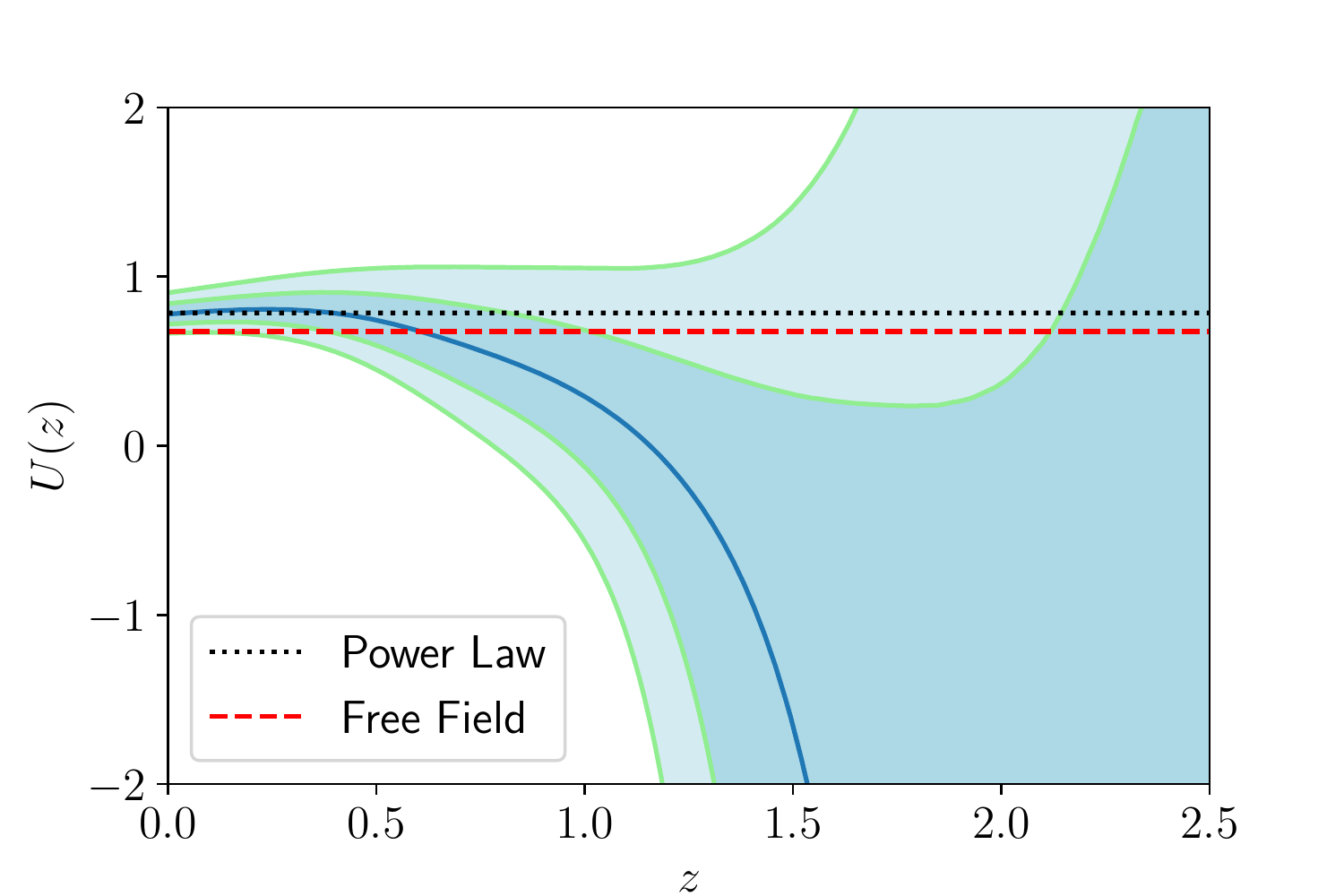}
    \caption{Results from the reconstructions from the large prior. {Left:} $U(z)$ reconstruction from 31 $H(z)$ data. {Right:} $U(z)$ reconstruction from Pantheon.}
    \label{PRPL}
\end{figure}

\begin{figure}
    \centering
    \includegraphics[width=.49\textwidth]{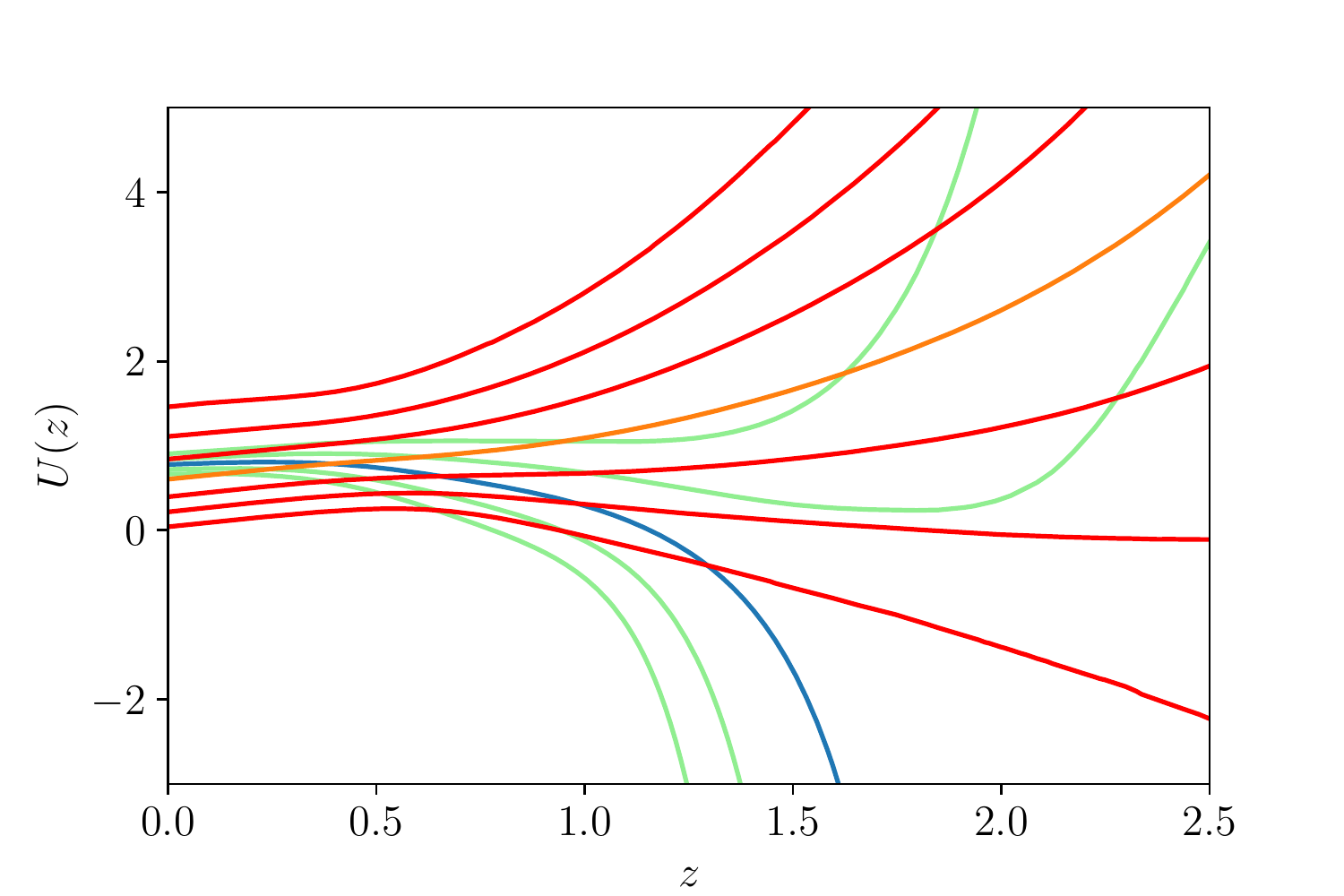}
    \includegraphics[width=.49\textwidth]{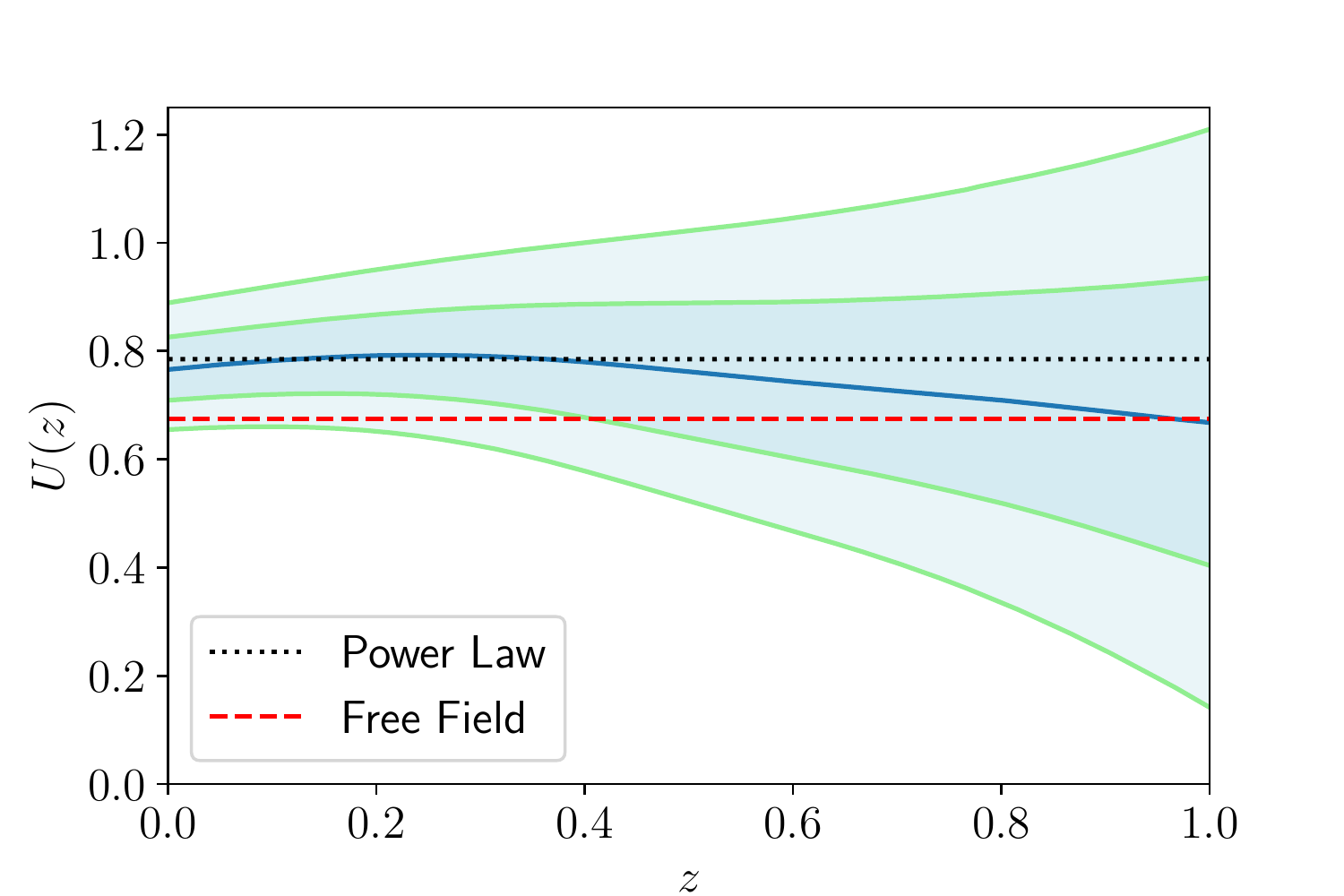}
    \caption{$U(z)$ reconstruction from $H(z)$ and Pantheon data {with a large prior over $\Omega_m$ and $\Omega_k$ (more details in the text)}. {Left:} Superposition of confidence intervals. SNe Ia reconstruction corresponds to red and $H(z)$ reconstruction corresponds to green continuous lines {Right:} Joint analysis.}
    \label{PCPL}
\end{figure}

The large prior reconstruction is presented in the Figures \ref{PRPL} and \ref{PCPL}.
This analysis is described similarly to the previous one. In Fig. \ref{PRPL}, the reconstruction of $U(z)$ shows each of the datasets separately: the curve on the left represents the reconstruction from {$H(z)$ data and the curve on the right represents the reconstruction of  SNe Ia Pantheon sample}. As in the previous analysis with Planck 18, we plotted the same evolution curves for the dimensionless potentials $U_{PL}(z)$ and $U_{FF}(z)$. { In the range $0< z <1.0$ the $U_{FF}(z)$ curve is compatible within $1\sigma$ with the $U(z)$ reconstruction, from $H(z)$ data, as we can see in the left panel of Fig. \eqref{PRPL}, the potential $U_{PL}(z)$ is within the $1\sigma$ region at approximately $55\%$ of the analyzed range.}

{Figure \ref{PRPL} (right) shows the reconstruction of $U(z)$ for Pantheon sample, where we can see that the $U_{FF}(z)$ curve does not exceed the $2 \sigma$ region of the reconstruction. The additional potential $U_{PL}(z)$ is compatible with the region of $1\sigma$ around $47\%$ of the redshift range. The results of the analysis for the large prior proved to be similar to what was obtained for the Planck prior, but now we find a better match between the reconstructions and the curves of both dimensionless test potentials, in most cases, due to increased regions of uncertainty.}

{In Figure \ref{PCPL} (left), we present the compilation of both reconstructions of $U(z)$ performed with a large prior. In this Figure, we can see that the confidence regions in 1$\sigma$, from both reconstructions, overlap for $z\lesssim 1.0$. In this redshift interval we obtain the reconstruction of $U(z)$ using the combination of observational data, as shown in Fig. \ref{PCPL} (right). Now the  $U_{PL}(z)$ curve is compatible within $1\sigma$ with the reconstruction inside this redshift interval and the $U_{FF}(z)$ curve is within the $1\sigma$ region of the reconstruction for $0.40\lesssim z<1$.}

{It is also worth to mention, as one can see on Figs. \ref{PRPP} and \ref{PRPL} that the potential is better constrained with SNe Ia data for lower redshifts while it is better constrained from $H(z)$ data at higher redshifts.}

{In order to show more quantitative results, in Table \ref{TUZ}, we show the $U(z =0)$ values with $1\sigma$ c.l. errors for the reconstructions of $U(z)$ via GP and also of $U_{PL}(z)$ and $U_{FF}(z)$ of the test models. We have found $1\sigma$ compatibility for the $U(z=0)$ values in all analyzed cases.}

%\begin{figure}
%    \centering
 %   \includegraphics[width=.8\textwidth]{new figs/PP_UCOM_CT.pdf}
  %  \caption{Joint analysis: $U(z)$ reconstruction from $H(z)$ and Pantheon data with a Planck 3$\sigma$ prior over $\Omega_m$ and $\Omega_k$.}
%    \label{PCPPJoint}
%\end{figure}

%\begin{figure}
 %   \centering
  %  \includegraphics[width=.8\textwidth]{new figs/LP_UCOMB_CT.pdf}
   % \caption{Joint analysis: $U(z)$ reconstruction from $H(z)$ and Pantheon data. Large prior.}
%    \label{PCLPJoint}
%\end{figure}

\begin{table}[H]
\begin{tabular}{|c|c|c|c|c|c|c|c|}
\hline
\multicolumn{2}{|c|}{ Model } &  \multicolumn{6}{|c|}{Gaussian Processes Reconstruction} \\\cline{1-8}
 \multirow{2}{*}{Power Law} & \multirow{2}{*}{Free Field} & \multicolumn{3}{|c|}{Planck Prior} & \multicolumn{3}{|c|}{Large Prior} \\\cline{3-8}
 & & $H(z)$ & SNe Ia & Joint & $H(z)$ & SNe Ia & Joint \\\hline$0.785^{+0.083}_{-0.063}$ & $0.675^{+0.113}_{-0.131}$&$0.611^{+0.219}_{-0.190}$ & $0.754^{+0.045}_{-0.044}$ & $0.747^{+0.044}_{-0.043}$ & $0.604^{+0.238}_{-0.208}$ & $0.778^{+0.062}_ {-0.059}$ & $0.766^{+0.060}_{-0.057}$ \\\hline
\end{tabular}
\caption{$U(z=0)$ from both scalar field models (Power Law and Free Field) and from both Gaussian Processes analysis, with different priors (Planck prior and Large prior), from $H(z)$, SNe Ia data and from the joint analysis.}
    \label{TUZ}
\end{table}

In \cite{Nair:2013sna}, the authors reconstruct the dark energy potential using SNe Ia and BAO, assuming a spatially flat universe. They find stronger constraints over $V(\phi)$, which can be explained by the fact they assume spatially flat universe and they use an older SNe Ia compilation, namely, Union 2.1, when supernova systematic errors were not as well understood as today. Their reconstruction is also restricted to lower redshifts, $z<1.4$ {Furthermore, they have not assumed the physical prior $\dot{\phi}^2>0$, which may lead to inconsistent results}.

A similar analysis was made for reconstruction of the Horndeski gravity \cite{Bernardo:2021qhu}, including the quintessence potential (sec. 4.1), where they use the Pantheon/MCT+BAO+CC $H(z)$ data, in the context of an spatially flat model and they find a behaviour for the scalar field potential similar to our Figs. \ref{PRPP} and \ref{PRPL}. They find more stringent limits to the scalar field potential due to the fact that they do not allow for the spatial curvature to vary.

\section{\label{conc}conclusion}
In this work, we have found constrains over the scalar field dark energy (SFDE, or quintessence) potential directly from $H(z)$ and SNe Ia data, without having to assume any form to the potential. We have also allowed for the spatial curvature and the matter parameter density to vary according two assumed priors, namely, a prior of 3$\sigma$ from CMB Planck 18 probe and a ``large prior'', which encompass many of the current observations.

As a result, we have found that two popular forms of the SFDE are compatible with our reconstructions, at least for some values of their parameters and some regions inside the data redshift interval. {Specifically, for $z=0$, we have shown that the test models are all compatible with the reconstructions within $1\sigma$ c.l.}

This analysis can be improved in the future when more data is available, mainly SNe Ia data at high redshift, as we have seen that the SNe Ia reconstruction have much higher errors at higher redshifts.

\appendix*
\section{\label{app}Particular Case \texpdf{$k = 0$}{k=0}}
An interesting case is the spatially flat cosmological model, with $k=0$, which is favoured in the context of the concordance model \cite{Aghanim:2018eyx}. In this context, the Friedmann equations \eqref{EF}-\eqref{EA} are now given by
%Assuming $k=0$ (spatial flatness), the equations are:
\begin{align}\label{FEQF}
H^2 &= \frac{8\pi G}{3}(\rho_m + \rho_\phi)\,,\\\label{FEA}
\frac{\ddot a}{a} &= -\frac{4\pi G}{3}\left(\rho_m +\rho_\phi + 3p_\phi\right)\,.
\end{align}
With the energy density of the field $\rho_{\phi}$ and its pressure $p_{\phi}$ given in [\ref{DEP}] the system \eqref{FEQF}-\eqref{FEA} can be written as
\begin{align}\label{FEF2}
    H^2 &= \frac{8\pi G}{3}\left[\rho_m +\frac{\dot\phi^2}{2}+V(\phi)\right]\,,\\\label{FEA2}
\frac{\ddot a}{a} &= -\frac{4\pi G}{3}\left[\rho_m +2
\dot\phi^2-2V(\phi)\right]\,.
\end{align}
So, from the equations \eqref{FEF2}-\eqref{FEA2} we can express the potential $V(\phi)$ as
\be 
V(\phi) = \frac{3H^2+\dot H}{8\pi G}-\frac{\rho_m}{2}\,.
\ee 
which matches the expression \eqref{EP} with the term $k=0$, as expected. Now, as we did in the previous section, let us write the relation for the potential $V(\phi)$ found above to the dimensionless potential $U(\phi)$ given by \eqref{PA}. Furthermore, using the relation \eqref{TZ}, we will change the time derivative to a derivative with respect to {\it redshift}.
\be
U(z)=E^2-\frac{E(1+z)}{3}\frac{dE}{dz}-\frac{\Omega_{m}(1+z)^3}{2}
\ee
We thus obtain the expression for the dimensionless potential $U(z)$ for a spatially flat universe. In this context, the transverse comoving distance $D_M(z)$ is reduced to the comoving distance $D_C(z)$ that is related to $E(z)$ through the relation \eqref{DCEz}. Deriving $E(z)$ with respect to {\it redshift} we obtain
\be
E'(z)=-\frac{D_C''(z)}{D_C'(z)^2}
\ee
Thus, we can write the dimensionless potential $U(z)$ in terms of the comoving distance $D_C(z)$ and its derivatives $D'_C(z)$ and $D''_C(z)$ as
\be
U(z)=\frac{3D_C'(z)+(1+z)D_C''(z)}{3D_C'(z)^3} - \frac{\Omega_{m}}{2}(1+z)^3\,.
\ee

\begin{acknowledgments}
RV is supported by  Funda\c{c}\~ao de Amparo \`a Pesquisa do Estado de S\~ao Paulo - FAPESP (thematic project process no. 2021/01089-1 and regular project process no. 2016/09831-0). SHP acknowledges financial support from  {Conselho Nacional de Desenvolvimento Cient\'ifico e Tecnol\'ogico} (CNPq)  (No. 303583/2018-5 and 308469/2021-6). This study was financed in part by the Coordena\c{c}\~ao de Aperfei\c{c}oamento de Pessoal de N\'ivel Superior - Brasil (CAPES) - Finance Code 001. {We thank Felipe Andrade-Oliveira for helpful discussions.}
\end{acknowledgments}

%\bibliographystyle{unsrtnat}
%\bibliographystyle{hunsrtnat}
%\bibliographystyle{h-physrev}

%\bibliography{references}

\end{document}